\documentclass[aip,amsmath,reprint]{revtex4-1}

\usepackage{graphicx}
\usepackage{subcaption}

\begin{document}


\title{On multichannel film dosimetry with channel-independent perturbations} 


\author{I. M{\'e}ndez}
\email[]{nmendez@onko-i.si}
\author{P. Peterlin}
\author{R. Hudej}
\author{A. Strojnik}
\author{B. Casar}
\affiliation{Department of Medical Physics, Institute of Oncology Ljubljana, Zalo\v{s}ka cesta 2, Ljubljana 1000, Slovenia}



\begin{abstract}
\textbf{Purpose:}
Different multichannel methods for film dosimetry have been proposed in the literature. Two of them are the weighted mean method and the method put forth by Micke \emph{et al} and Mayer \emph{et al}. The purpose of this work was to compare their results and to develop a generalized channel-independent perturbations framework in which both methods enter as special cases.

\noindent\textbf{Methods:}
Four models of channel-independent perturbations were compared: weighted mean, Micke-Mayer method, uniform distribution and truncated normal distribution. A closed-form formula to calculate film doses and the associated Type B uncertainty for all four models was deduced.

To evaluate the models, film dose distributions were compared with planned and measured dose distributions. At the same time, several elements of the dosimetry process were compared: film type EBT2 versus EBT3, different waiting-time windows, reflection mode versus transmission mode scanning, and planned versus measured dose distribution for film calibration and for $\gamma$-index analysis.

The methods and the models described in this study are publicly accessible through IRISEU. Alpha 1.1 (http://www.iriseu.com). IRISEU. is a cloud computing web application for calibration and dosimetry of radiochromic films.

\noindent\textbf{Results:}
The truncated normal distribution model provided the best agreement between film and reference doses, both for calibration and $\gamma$-index verification, and proved itself superior to both the weighted mean model, which neglects correlations between the channels, and the Micke-Mayer model, whose accuracy depends on the properties of the sensitometric curves. 

With respect to the selection of dosimetry protocol, no significant differences were found between transmission and reflection mode scanning, between $75 \pm 5$ min and $20 \pm 1$ h waiting-time windows or between employing EBT2 or EBT3 films. Significantly better results were obtained when a measured dose distribution was used instead of a planned one as reference for the calibration, and when a planned dose distribution was used instead of a measured one as evaluation for the $\gamma$-analysis.

\noindent\textbf{Conclusions:}
The truncated normal distribution model of channel-independent perturbations was found superior to the other three models under comparison and we propose its use for multichannel dosimetry.

\end{abstract}


\maketitle 

\section{Introduction}

Radiochromic film dosimetry with flatbed scanners and Gafchromic films (Ashland Inc., Wayne, NJ) has been extensively studied in the literature\cite{devic:2005, Fuss:2007, Paelinck:2007, battum:2008, Martisikova:2008, Richley:2010, Devic:2011}. High spatial resolution, near water equivalence\cite{crijns:2102, aapm:55} and weak energy dependence\cite{rink:2007, richter:2009, arjomandy:2010, lindsay:2010, massillon} make radiochromic films convenient for measurements whenever sharp dose gradients, tissue heterogeneities or charged particle disequilibrium conditions exist. This opens up a wide range of applications for radiochromic films in the field of radiotherapy.

Recently, different multichannel dosimetry methods have been proposed to take into account the information conveyed by all three color channels delivered by the scanner. Micke \emph{et al}\cite{AMicke:2011} proposed the use of channel-independent perturbations to compensate for variations in the thickness of the active layer, artifacts, nonuniform response of the scanner or other disturbances. They found a substantial gain in dosimetric accuracy using this method. Van Hoof \emph{et al}\cite{vanHoof:2012} found that this method performs at least as well as the conventional single-red-channel dosimetry. Mayer \emph{et al}\cite{mayer:2012} derived a closed-form solution to obtain the dose employing channel-independent perturbations. They also compared different single, dual and triple channel methods, and found better agreement between planned and calculated dose distributions using the average dose of all three channels in comparison to using the channel-independent perturbations method. In an earlier article\cite{mendez:2013}, our group suggested calculating the film dose as the weighted mean dose of all three channels. For each channel, the inverse of the mean square error obtained during the film calibration was used as weight. With this method, triple-channel dosimetry was found to be substantially superior to single-red-channel dosimetry. 

The purpose of this work is to compare both weighted mean and Micke-Mayer methods, considering them as special cases of a more general channel-independent perturbations method. Deficiencies and important problems associated with both methods will be explained. To overcome these problems, an improved multichannel film dosimetry method will be introduced. Its performance against the other methods will be verified by comparing film dose distributions with planned as well as with measured dose distributions. In addition, other elements of the dosimetry process will be compared: film types\cite{reinhardt:2012}, scanning modes\cite{park:2012}, scanning waiting-time windows\cite{devic:2010} and choices of reference dose distribution. 

\section{Methods and materials}

\subsection{Channel-independent perturbations}
Channel-independent perturbations are obtained by applying a first order Taylor expansion to the dose due to a small perturbation:
 
\begin{equation}
\label{eqsystem}
\begin{cases}
D(r) = D_{R}(r) + \dot{D}_{R}(r) \Delta(r) + \epsilon_{R}(r)\\
D(r) = D_{G}(r) + \dot{D}_{G}(r) \Delta(r) + \epsilon_{G}(r)\\
D(r) = D_{B}(r) + \dot{D}_{B}(r) \Delta(r) + \epsilon_{B}(r)
\end{cases},
\end{equation}

Micke \emph{et al}\cite{AMicke:2011} derived the dose from the optical density (OD) of the irradiated film. Mayer \emph{et al}\cite{mayer:2012} used pixel values directly. In this study, better results were found in preliminary tests using net optical density\citep{devic:2005} (NOD) in comparison to using OD. Therefore, the channel-independent perturbation consists of a change in NOD and is represented by $\Delta(r)$. $D(r)$ represents the dose absorbed by the film at point $r$. $D_{k}$ is the absolute dose measured by the channel $k$, {\it i.e.}, red (R), green (G) or blue (B) channel, when no disturbance is present, and it is calculated directly from the calibration model. In this study, the calibration model includes the lateral correction\cite{Paelinck:2007, fiandra:2006, battum:2008, devic:2006, lynch:2006, Martisikova:2008, Fuss:2007, menegotti:2008, saur:2008} and the sensitometric curve. $\dot{D}_{k}(r)$ is the first derivative of the dose, with respect to the NOD, at point $r$. Finally, $\epsilon_{k}(r)$ is an error term accounting for the difference between the dose absorbed by the film and the dose measured in the channel $k$ after correction by the perturbation. 

Both for reflection\cite{ohuchi:2007} and transmission mode scanning, the NOD, denoted by $z$, was defined as

\begin{equation}
z := \log_{10} \frac{v_{\mathrm{nonirr}}}{v_{\mathrm{irr}}},
\end{equation}

where $v_{\rm{nonirr}}$ and $v_{\rm{irr}}$ represent pixel values of nonirradiated and irradiated films, respectively, after applying lateral corrections. Our previous results\cite{mendez:2013} found better fit when lateral corrections are absolute corrections independent of dose, and sensitometric curves are polynomial fits of order four. Hence, lateral corrections were calculated as

\begin{equation}
\label{lateral}
v_{k} = a_{k_1} (x-x_{c}) + a_{k_2} (x-x_{c})^{2} + \hat{v_{k}}, 
\end{equation}

where $\hat{v_{k}}$ represents uncorrected pixel values, the $x$ axis is parallel to the CCD array, $x_c$ is the {\it x} coordinate of the center of the scanner, $v_{k}$ represents corrected pixel values, and $a_{k}$ are fitting parameters. Sensitometric curves followed

\begin{equation}
\label{sensitometric}
D_{k} = \sum_{j=1}^{4} b_{k_j} z_{k}^j ,
\end{equation}

and $\dot{D}_{k}$ was 

\begin{equation}
\dot{D}_{k} = \sum_{j=1}^{4} j b_{k_j} z_{k}^{j-1} ,
\end{equation}

where $b_{k}$ are fitting parameters.

\subsection{Solving the equation system}
The values of $\Delta(r)$ and $\epsilon_{k}(r)$ for $k=R,G,B$ in Eq.(\ref{eqsystem}) are unknown. As a result, the absorbed dose $D(r)$ cannot be obtained directly. However, one can examine different probability density functions (pdf) for $\Delta(r)$ and $\epsilon_{k}(r)$ and, if $D(r)$ is known, analyze how well these models reproduce the absorbed dose distribution.

\subsubsection{Probability density function of the dose}
Given the pdfs of $\Delta$, symbolized by $f(\Delta)$, and of each $\epsilon_{k}$, symbolized by $g_{k}(\epsilon_{k})$, the joint pdf of $D$, symbolized by $P(D)$, is:

\begin{equation}
P(D) = \int  f(\Delta) \prod_{k} g_{k}(D - D_{k} - \dot{D}_{k} \Delta)  \:\mathrm{d}\Delta , 
\end{equation}

taking into account that $\Delta$ and $\epsilon_{k}$ are not independent from each other:
\begin{equation}
\epsilon_{k} = D - D_{k} - \dot{D}_{k} \Delta .
\end{equation}

Let us consider that the error terms are distributed normally with zero mean and $\sigma_{k}^{2}$ variance: 
\begin{equation}
g_{k}(\epsilon_{k}) = \mathcal{N}(0,\sigma_{k}^{2})
\end{equation}

The joint pdf of $D$ becomes:
\begin{equation}
P(D) = \int  f(\Delta) \prod_{k} \frac{1}{\sigma_{k} \sqrt{2\pi}} \:\mathrm{e}^{-\frac{1}{2} \left(\frac{D - D_{k} - \dot{D}_{k}\Delta}{\sigma_{k}}\right)^{2}}  \:\mathrm{d}\Delta.
\end{equation}
   
Three different models for $f(\Delta)$ will be considered: 

a) Normally distributed perturbation ({\it i.e.}, $f(\Delta) = \mathcal{N}(0,\sigma_{\Delta}^{2})$):

\begin{equation}
P(D) =  \frac{1}{(2\pi)^{\frac{n}{2}} \sigma_{\Delta} \prod_{k=1}^{n} \sigma_{k}} \: \frac{1}{\sqrt{A}} \: \:\mathrm{e}^{-\frac{1}{2} \left(C - \frac{B^2}{4A}\right)}, 
\end{equation}

where $n$ represents the number of color channels ({\it i.e.}, $n = 3$) and

\begin{equation}
\label{Adef}
A= \frac{1}{\sigma_{\Delta}^{2}} + \sum_{k} \left( \frac{\dot{D}_{k}}{\sigma_{k}} \right)^{2}
\end{equation}

\begin{equation}
B=-2\sum_{k} \frac{(D - D_{k})\dot{D}_{k}}{\sigma_{k}^{2}}
\end{equation}

\begin{equation}
C=\sum_{k} \left( \frac{D - D_{k}}{\sigma_{k}} \right)^{2}.
\end{equation}

b) Truncated normal distribution with $\Delta \in (-\theta, \theta)$:

\begin{equation}
\label{truncated}
P(D)\propto  \:\mathrm{e}^{-\frac{1}{2} \left(C - \frac{B^2}{4A}\right)} \: \left( \mathrm{erf} \left(\frac{\theta + \frac{B}{2A}}{\sqrt{\frac{2}{A}}}\right) - \mathrm{erf} \left(\frac{-\theta + \frac{B}{2A}}{\sqrt{\frac{2}{A}}}\right)\right) , 
\end{equation}

excluding a normalizing term independent of $D$.

c) Uniform distribution with $\Delta \in (-\theta, \theta)$: is a special case of Eq.(\ref{truncated}) where $\sigma_{\Delta}$ goes to infinity, therefore $A= \sum_{k} \left( \frac{\dot{D}_{k}}{\sigma_{k}} \right)^{2}$.

\subsubsection{Dose calculation}
The most likely value of the absorbed dose $D$, symbolized by $d$, is the one that maximizes $P(D)$. The exponential term in $P(D)$, $P(D) \propto \:\mathrm{e}^{-\frac{1}{2} \left(C - \frac{B^2}{4A}\right)}$, can be expressed in terms of $D$ as a gaussian function:

\begin{equation}
P(D) \propto \:\mathrm{e}^{-\frac{1}{2} \left(\frac{D - \mu_{D}}{\sigma_{D}}\right)^{2}}, 
\end{equation}

where 

\begin{equation}
\label{Dose}
\mu_{D} = d = \frac{A\beta - \gamma\delta}{A\alpha-\gamma^{2}} 
\end{equation}

and 

\begin{equation}
\label{uncertainty}
\sigma_{D} = \sqrt{\frac{A}{A\alpha-\gamma^{2}}}, 
\end{equation}

A is defined in Eq.(\ref{Adef}) and

\begin{equation}
\alpha=\sum_{k} \frac{1}{\sigma_{k}^{2}}
\end{equation}

\begin{equation}
\beta=\sum_{k} \frac{D_{k}}{\sigma_{k}^{2}}
\end{equation}

\begin{equation}
\gamma=\sum_{k} \frac{\dot{D}_{k}}{\sigma_{k}^{2}}
\end{equation}

\begin{equation}
\delta=\sum_{k} \frac{D_{k}\dot{D}_{k}}{\sigma_{k}^{2}} .
\end{equation}

Eq.(\ref{Dose}) and Eq.(\ref{uncertainty}) can be considered, respectively, as the estimated absolute dose and its type B uncertainty \cite{JCGM:2008}. This result is exact for normally distributed perturbations and an approximation for truncated normal and uniform distributions.

\subsection{Models of channel-independent perturbations under comparison}

Four models of channel-independent perturbations were compared: weighted mean (WM), Micke-Mayer (MM) method, uniform distribution (UD) and truncated normal distribution (TN). They are summarized in Table~\ref{tab:models}.

The weighted mean method is a limit case of Eq.(\ref{eqsystem}) in which $\Delta(r) = 0$. Thus, all three channels are independent of each other, which implies that correlations between channels are neglected.

The method employed by Micke \emph{et al}\cite{AMicke:2011} and Mayer \emph{et al}\cite{mayer:2012} is a special case of Eq.(\ref{eqsystem}) where all $\sigma_{k}$ are equal and $f(\Delta)$ is uniformly distributed. Under these premises, Eq.(\ref{Dose}) becomes:

\begin{equation}
d = \frac{A\beta - \gamma\delta}{A\alpha-\gamma^{2}} = \frac{(\sum_{k=1}^{n} \dot{D}_{k}) (\sum_{k=1}^{n} D_{k} \dot{D}_{k}) - (\sum_{k=1}^{n} \dot{D}_{k}^{2}) (\sum_{k=1}^{n} D_{k})}{(\sum_{k=1}^{n} \dot{D}_{k})^{2} - n (\sum_{k=1}^{n} \dot{D}_{k}^{2})},
\end{equation}

which coincides with the closed-form solution derived by Mayer \emph{et al} \cite{mayer:2012}. The uncertainty in the dose associated to this model becomes:

\begin{equation}
\label{MMuncertainty}
\sigma_{D} = \sqrt{\frac{A}{A\alpha-\gamma^{2}}} = \sigma_{k} 
\sqrt{\frac{\sum_{k=1}^{n} \dot{D}^{2}_{k}}
{n \sum_{k=1}^{n} \dot{D}^{2}_{k} - (\sum_{k=1}^{n} \dot{D}_{k})^2}} 
\end{equation}

The uniform distribution model is a more general and realistic model for the perturbation than the MM one. In this case, $f(\Delta)$ is uniformly distributed but the $\sigma_{k}$ can differ.

Finally, the truncated normal distribution model considers that $f(\Delta)$ follows a truncated normal distribution. The WM model is a limit case and the UD and MM models are particular cases of this model.  

\begin{table*}
\caption{\label{tab:models} Models of channel-independent perturbations under comparison.}

\begin{ruledtabular}
\begin{tabular}{lcc}
Model & Abbreviation & Assumptions \\
\hline
Weighted mean & WM & $\Delta(r) = 0$\\
Micke-Mayer method & MM & $f(\Delta)$ uniform distribution, $\sigma_{k}$ are equal\\
Uniform distribution & UD & $f(\Delta)$ uniform distribution\\
Truncated normal distribution & TD & $f(\Delta)$ truncated normal distribution\\
\end{tabular}
\end{ruledtabular}
\end{table*}
 
\subsection{Measurement protocol}
Ten 8 ${\rm inch}$ $\times$ 10 ${\rm inch}$ EBT2 films from lot A03171101A and seventeen EBT3 films from lot A05151201 were employed. They were handled following recommendations from the AAPM TG-55 report \cite{aapm:55}. 

Films were scanned with an Epson Expression 10000XL flatbed scanner (Seiko Epson Corporation, Nagano, Japan) using Epson Scan v.3.0 software. Images were acquired in 48-bit RGB mode (16 bit per channel), the resolution was 72 dpi (0.35 mm/px) and the image processing tools were turned off. 

Before acquisitions, the scanner was warmed up for at least 30 min. After the warm-up, and whenever long interruptions occurred, five empty scans were taken to stabilize the temperature of the scanner lamp. Films were centered on the scanner with a black opaque cardboard frame and scanned in portrait orientation. Five consecutive scans were made for each film. To avoid the warm-up effect of the lamp due to multiple scans\cite{Paelinck:2007, Martisikova:2008} the first scan was discarded and the resulting image was the average of the remaining four. 

Films were scanned before irradiation both in reflection and in transmission mode. After irradiation, two waiting-time windows were studied: films were first scanned after $75 \pm 5$ min in transmission mode, and again after $20 \pm 1$ h both in reflection and transmission mode.
   
Irradiation was delivered with a 6 MV photon beam from a Novalis Tx accelerator (Varian, Palo Alto, CA, USA). Three different phantoms were used: CIRS Thorax Phantom (Model 002LFC, Computerized Imaging Reference Systems Inc. Norfolk, VA, USA), CIRS Pelvic Phantom (Model 002PRA) and IBA MatriXX Evolution MULTICube (IBA Dosimetry GmbH, Germany). Source-axis distance (SAD) setup was used for all three phantoms. To avoid the films lying in the beam axis plane\cite{Kunzler:2009}, the films were placed at an offset of 1.5 cm from the beam axis in the CIRS Thorax Phantom  and of 1.3 cm in the CIRS Pelvic Phantom. The IBA MatriXX Evolution MULTICube was used jointly with the IBA MatriXX Evolution ionization chamber array, which measured the dose distribution delivered. The film was situated atop the detector.

The absolute dose distributions in the plane of the film were calculated with Eclipse v.10.0 (Varian Medical Systems, Palo Alto, CA, USA) treatment planning system (TPS) using the anisotropic analytical algorithm (AAA). The planned dose distributions were exported to dose matrices with a resolution of 0.49 mm/px. The dose values were scaled to correct for the daily output of the linac. Whenever MatriXX Evolution was used, the dose distribution was simultaneously measured. The dose values were scaled with a constant factor to correct for the distance (which was 3.5 mm) between the film and the plane at the effective depth of measurement. The MatriXX 2D array has a resolution of 7.62 mm/px. Planned and measured dose distributions were bicubically interpolated to the resolution of the scan and registered with the film.  

Film scans, planned dose ditributions and measured dose distributions were uploaded and processed with IRISEU. Alpha 1.1 (http://www.iriseu.com). IRISEU. is a cloud computing web application for calibration and dosimetry of radiochromic films. It is developed by one of the authors (IM) and incorporates the methods and models described in this study. It was employed for the calibration, dosimetry and gamma index evaluation. Additional statistical analysis was performed with R statistical software\cite{R:software}.

\subsection{Calibration}

To fit the calibration parameters, the plan-based method\cite{mendez:2013} was chosen. Besides being faster than the calibration method with fragments, the plan-based method provides a more representative sample of perturbations (since it can use every pixel of the film). This method requires one or more 2D dose distributions as reference doses for the calibration. In order to obtain them, films were placed in the MatriXX Evolution phantom and irradiated with a $60^{\circ}$ Enhanced Dynamic Wedge (EDW) field of dimensions 20$\times$20 ${\rm cm^2}$. To reduce intralot variations\cite{mendez:2013}, three separate films from each lot were exposed. The range of doses relevant for this study and for posterior clinical use was estimated between 50 cGy and 400 cGy. To encompass the whole range, two different fields were used: the wedge dose spanned from approximately 75 cGy to approximately 400 cGy (535 MU) for two of the films from each lot and from approximately 50 cGy to approximately 300 cGy (401 MU) for the remaining one. 

Following this procedure, one set with EBT2 and another with EBT3 films were irradiated. Posteriorly, the films were scanned following the three protocols previously mentioned: reflection mode with $20 \pm 1$ h time window, transmission mode with $20 \pm 1$ h time window and transmission mode with $75 \pm 5$ min time window. Each set of images (six sets in total) was employed to calibrate each of the four models of channel-independent perturbations. Each of the models was calibrated against planned dose distributions (calculated with the TPS) and against measured dose distributions (simultaneously measured with MatriXX during the irradiations). Altogether, a total of 48 calibrations were computed.

Pixel values of the films exposed were translated into doses, for each color channel independently, fitting the calibration parameters. A genetic algorithm was used to fit the parameters minimizing the root-mean-square error (RMSE) of the differences between film doses for each channel ($D_{k}(r)$) and reference doses ($D(r)$).

This optimization provides the parameters used in Eq.(\ref{lateral}) and Eq.(\ref{sensitometric}). This is enough for film dosimetry following WM or MM models. However, to obtain $d(r)$ using UD or TN models $\sigma_{k}$ are necessary, and also $\sigma_{\Delta}$ if using the TN model. Knowing lateral correction, sensitometric curve parameters and the standard deviation of $f(\Delta)$, which depends on $\sigma_{\Delta}$ and $\theta$ and will be symbolized by $\tilde{\sigma}_{\Delta}$, $\sigma_{k}$ can be estimated with

\begin{equation}
\label{eq_cal}
\tilde{\sigma}_{k}^{2} \simeq (E[\dot{D}_{k}] \: \tilde{\sigma}_{\Delta})^{2} + \sigma_{k}^{2} 
\end{equation}

where $\tilde{\sigma}_{k}$ is the RMSE of the channel and $E[\dot{D}_{k}]$ is the expected value ({\it i.e.}, mean) of $\dot{D}_{k}$. 

The values of $\tilde{\sigma}_{\Delta}$ for UD and TN models, and of $\sigma_{\Delta}$ for the TN model, were obtained optimizing the RMSE of the differences between film doses ($d(r)$) and reference doses ($D(r)$). 
  
\subsection{Verification}
To evaluate the four models of channel-independent perturbations, film dose distributions were compared with planned and with measured dose distributions. Global gamma analysis was conducted. The tolerances were 4 \%, 3 mm with 20\% of the maximum dose as threshold. Fourteen different cases were tested (Table~\ref{tab:Tests}). The cases were chosen with the intention of compiling a representative sample of dose distributions: several simple geometries, tissue heterogeneities, three-dimensional conformal radiotherapy (3D-CRT) plans and  intensity modulated radiation therapy (IMRT) plans, including volumetric modulated arc therapy (VMAT) plans, were selected. EBT3 films were irradiated with all the cases but only a subset (considered representative) of them was used with EBT2 films, as shown in Table~\ref{tab:Tests}. Appropriate phantoms were employed dependent on the test case.

At the same time, several elements of the dosimetry process were compared: film type EBT2 versus EBT3, different waiting-time windows ({\it i.e.}, $75 \pm 5$ min versus $20 \pm 1$ h), reflection versus transmission mode scanning and planned versus measured reference dose distribution for film calibration and for gamma index analysis.

As a result, seven EBT2 and fourteen EBT3 films were irradiated with the cases shown in Table~\ref{tab:Tests}. They were scanned following the three scanning protocols under study. Each image was translated into a dose distribution following each of the four models of channel-independent perturbations. The film dose distributions were compared with the planned dose distributions in the plane of the film. Whenever the test was irradiated in the MatriXX phantom, the film dose distributions were also compared with the measured dose distributions. When film dose distributions were compared with planned ones, the calibration parameters of the model had been fitted using planned reference dose distributions, and analogously with measured dose distributions. If both planned and measured reference dose distributions are accurate, they should provide similar sets of calibration parameters. Following this, and for the TN model only, film distributions obtained with calibration parameters fitted using measured reference dose distribution were also compared with planned dose distributions, and vice versa ({\it i.e.}, film distributions obtained with calibration parameters fitted using planned reference dose distribution were compared with dose distributions measured with MatriXX).

\begin{table*}
\caption{\label{tab:Tests} Description of the test cases, including film type and phantom used in the measurements.}
\begin{ruledtabular}
\begin{tabular}{lccr}
Test & Description & Film type & Phantom \\
\hline
A & Square 15$\times$15 ${\rm cm^2}$ & EBT2,EBT3 & MatriXX\\ 
B & Chair test \citep{vanesch:2002}& EBT2,EBT3 & MatriXX\\ 
C & Pyramid shaped in both axis\citep{ju:2002}& EBT2,EBT3 & MatriXX\\ 
D & EDW $30^{\circ}$ field & EBT3 & MatriXX\\
E & EDW $45^{\circ}$ collimator 90 field & EBT3 & MatriXX\\ 
F & Y-shaped 3D CRT field & EBT2,EBT3 & MatriXX\\ 
G & Predominantly convex shaped 3D CRT field & EBT3 & MatriXX \\ 
H & RapidArc prostate 1 & EBT3 & CIRS Pelvic \\  
I & RapidArc prostate 2 & EBT3 & CIRS Pelvic \\  
J & RapidArc prostate 3 & EBT2,EBT3 & CIRS Pelvic \\  
K & Square 10$\times$10 ${\rm cm^2}$, lung inhomogeneity  & EBT3 & CIRS Thorax \\ 
L & Lateral incidence, lung inhomogeneity  & EBT2,EBT3 & CIRS Thorax \\ 
M & Four field box, lung inhomogeneity  & EBT3 & CIRS Thorax \\ 
N & EDW and asymmetric fields, lung inhomogeneity  & EBT2,EBT3 & CIRS Thorax \\ 
\end{tabular}
\end{ruledtabular}
\end{table*}

\section{Results and discussion}

Twenty-four different dosimetry protocols were analyzed in this study. To represent each protocol in a clear and concise way, they will be named using four characters (Table~\ref{tab:symbols}). The characters stand for: gamma analysis with either planned (P) or measured (M) evaluation dose distributions, scanning in reflection mode with $20 \pm 1$ h time window (R), in transmission mode with $20 \pm 1$ h time window (T) or in transmission mode with $75 \pm 5$ min time window (t), film type either EBT2 (2) or EBT3 (3) and calibration with either planned (p) or measured (m) reference dose distributions.

\begin{table*}
\caption{\label{tab:symbols} Elements of the dosimetry protocol under comparison.}

\begin{ruledtabular}
\begin{tabular}{lcc}
Element of the protocol & Alternative & Abbreviation \\
\hline
Evaluation dose distribution for the gamma analysis & Planned & P\\
 & Measured & M\\
Scanning mode and time window & Reflection, $20 \pm 1$ h & R\\
 & Transmission, $20 \pm 1$ h & T\\
 & Transmission, $75 \pm 5$ min & t\\
Film type & EBT2 & 2\\
 & EBT3 & 3\\
Reference dose distribution for the calibration & Planned & p\\
 & Measured & m\\
\end{tabular}
\end{ruledtabular}
\end{table*}

\subsection{Selection of model of channel-independent perturbations}

\begin{table*}
\caption{\label{tab:Main} Comparison of film doses ($d(r)$) with planned or measured doses ($D(r)$). Data are aggregated by model of channel-independent perturbations and dosimetry protocol. It contains RMSEs from the calibrations as well as gamma mean ($\overline{\gamma}$) and percentage of points with $\gamma_{<1}$ from the verification gamma analysis. The models of channel-independent perturbations include: weighted mean (WM), Micke-Mayer (MM) method, uniform distribution (UD) and truncated normal distribution (TN).}
\begin{ruledtabular}
\begin{tabular}{l ccc ccc ccc ccc}
 & \multicolumn{3}{c}{WM} & \multicolumn{3}{c}{MM} & \multicolumn{3}{c}{UD} & \multicolumn{3}{c}{TN}\\
\cline{2-4} 
\cline{5-7}
\cline{8-10}
\cline{11-13}

& RMSE & $\overline{\gamma}$ & $\gamma_{<1}$ & RMSE & $\overline{\gamma}$ & $\gamma_{<1}$ & RMSE & $\overline{\gamma}$ & $\gamma_{<1}$ & RMSE & $\overline{\gamma}$ & $\gamma_{<1}$\\
Protocol  & (cGy) & & (\%) & (cGy) & & (\%) & (cGy) & & (\%) & (cGy) & & (\%)\\
\hline
PR2p & 3.5 & 0.18 & 98.4 & 5.4 & 0.28 & 93.9 & 5.2 & 0.27 & 93.6 & 2.8 & 0.19 & 98.0 \\
Pt2p & 3.8 & 0.23 & 96.9 & 3.3 & 0.22 & 96.9 & 3.2 & 0.22 & 97.2 & 2.8 & 0.19 & 98.0 \\
PT2p & 5.1 & 0.32 & 93.2 & 3.7 & 0.22 & 98.1 & 3.5 & 0.22 & 97.5 & 3.1 & 0.21 & 97.1 \\
PR3p & 2.9 & 0.18 & 98.1 & 4.9 & 0.26 & 95.1 & 4.8 & 0.25 & 95.6 & 2.7 & 0.15 & 98.7 \\
Pt3p & 6.2 & 0.24 & 96.3 & 4.1 & 0.19 & 98.6 & 4.1 & 0.19 & 98.6 & 4.1 & 0.17 & 99.4 \\
PT3p & 3.3 & 0.21 & 97.1 & 3.6 & 0.17 & 99.1 & 3.3 & 0.14 & 99.2 & 2.6 & 0.14 & 99.4 \\
MR2m & 3.6 & 0.31 & 98.1 & 5.1 & 0.39 & 93.1 & 4.9 & 0.39 & 92.6 & 2.9 & 0.31 & 97.3 \\
Mt2m & 4.4 & 0.35 & 94.6 & 3.2 & 0.34 & 95.3 & 3.1 & 0.33 & 96.1 & 2.9 & 0.32 & 96.7 \\
MT2m & 5.4 & 0.36 & 91.5 & 3.5 & 0.31 & 97.3 & 3.2 & 0.35 & 96.8 & 3.2 & 0.33 & 97.5 \\
MR3m & 3.1 & 0.28 & 97.7 & 5.0 & 0.44 & 88.0 & 4.8 & 0.29 & 98.1 & 2.9 & 0.29 & 98.1 \\
Mt3m & 6.2 & 0.34 & 94.2 & 4.2 & 0.30 & 96.9 & 4.2 & 0.28 & 97.8 & 4.2 & 0.28 & 97.8 \\
MT3m & 4.1 & 0.38 & 92.1 & 5.1 & 0.28 & 97.1 & 4.8 & 0.28 & 97.2 & 3.8 & 0.32 & 95.8 \\
\end{tabular}
\end{ruledtabular}
\end{table*}

Table~\ref{tab:Main} compares film doses ($d(r)$) with planned or measured doses ($D(r)$), data are aggregated by model of channel-independent perturbations and dosimetry protocol. It contains RMSEs from the calibrations as well as gamma mean ($\overline{\gamma}$) and percentage of points with $\gamma_{<1}$ from the verification gamma analysis.

\begin{figure*}
\begin{minipage}[b]{0.47\linewidth}
\centering
\includegraphics[width=\linewidth]{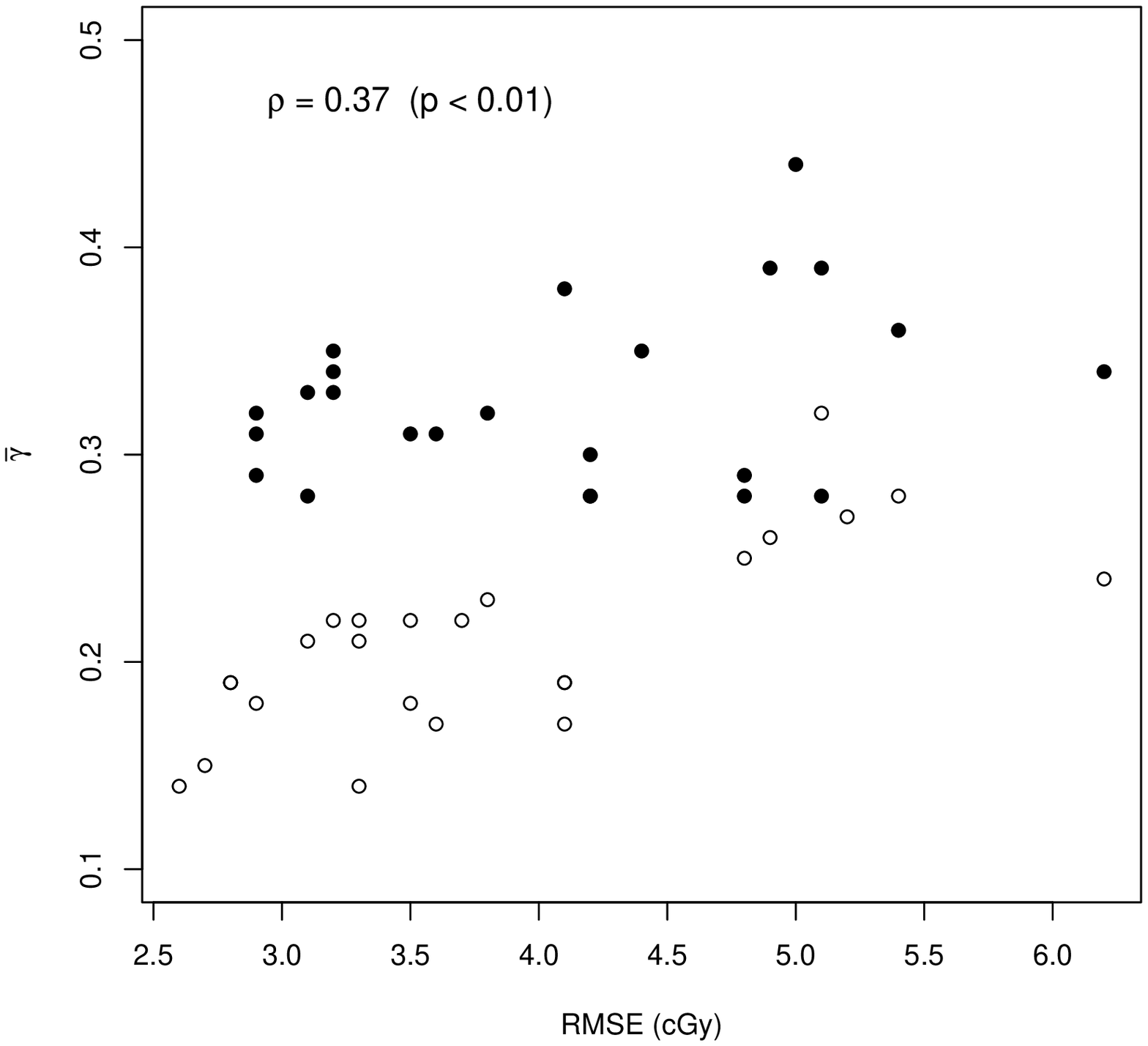}\\
(a)
\end{minipage}
\hfill
\begin{minipage}[b]{0.47\linewidth}
\centering
\includegraphics[width=\linewidth]{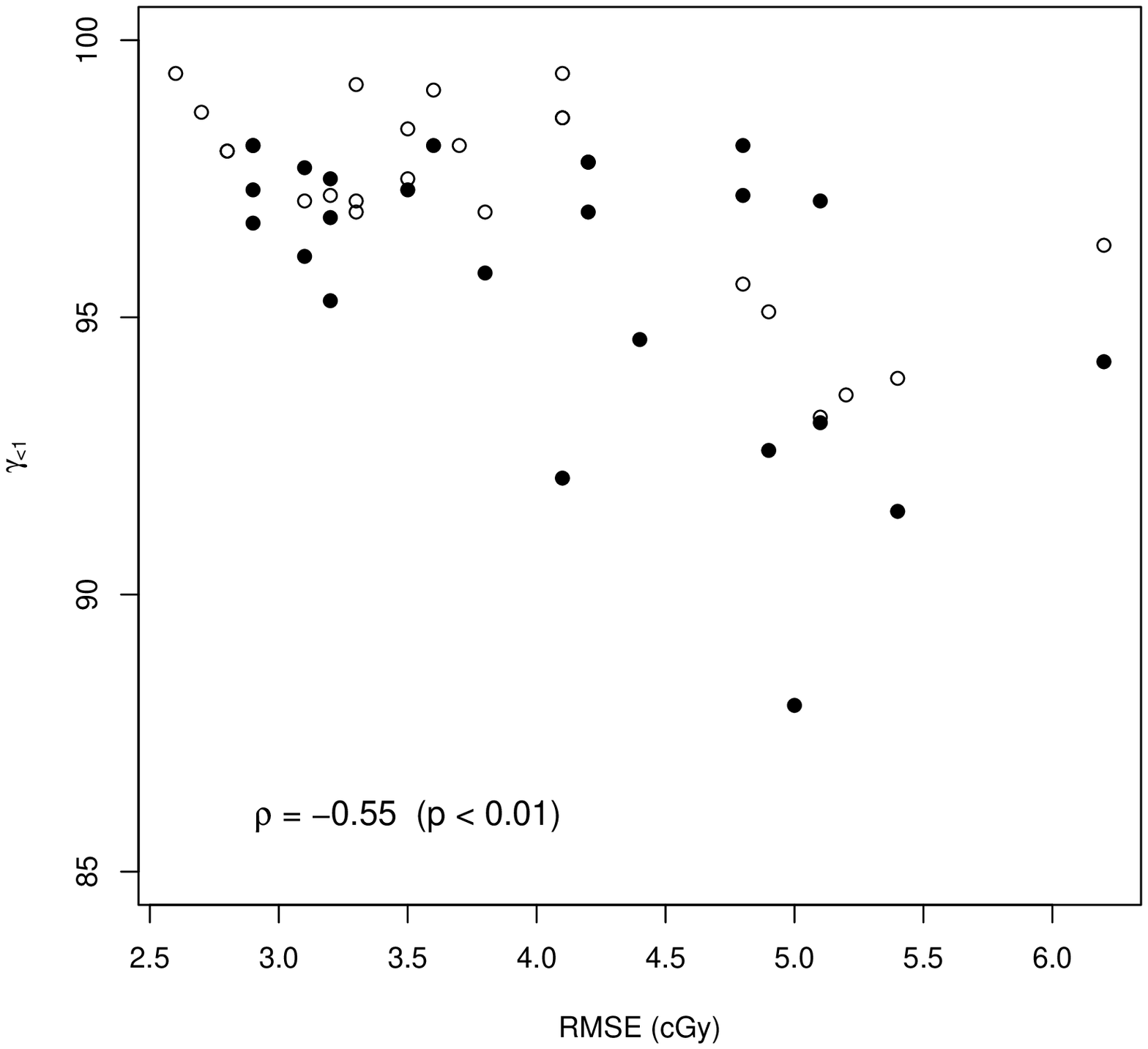}\\
(b)
\end{minipage}
\caption{\label{fig:correlation} Correlation between calibration RMSEs and (a) $\overline{\gamma}$ or (b) $\gamma_{<1}$, from Table~\ref{tab:Main}. White dots correspond to planned dose distributions and black dots to measured dose distributions. Spearman's rank correlation coefficients and $p$-values of the correlations are shown.}
\end{figure*}

Considering the size of the sample and calculating likelihood from RMSE, Akaike Information Criterion (AIC) values are equivalent to calibration RMSEs. Since the TN model provided as good or better calibration RMSEs than the other models in all protocols, according to the AIC the TN model should be selected for multichannel dosimetry.

\begin{table*}
\caption{\label{tab:Pair_models} Comparison of models employing paired difference tests. Models are paired for each dosimetry protocol. Differences are in $\overline{\gamma}$ values from Table~\ref{tab:Main}. Mean differences between models are shown. Negative values indicate that the first model obtained better results than the second one, and the opposite for positive values. Between parentheses, $p$-values of the t-tests are shown. The null hypothesis is that the mean difference between paired observations is zero. Protocols are grouped into protocols R and T.}     

\begin{ruledtabular}
\begin{tabular}{lcccccc}
Protocol group & WM - MM & WM - UD & WM - TN & MM - UD & MM - TN & UD - TN\\
\hline
R + T & -0.03 (0.40) & 0.01 (0.61) & 0.03 (\textless 0.05) & 0.02 (0.26) & 0.06 (\textless 0.05) & 0.04 (\textless 0.05) \\
R & -0.11 (\textless 0.05) & -0.03 (0.51) & 0.00 (0.81) & 0.04 (0.32) & 0.11 (\textless 0.05) & 0.07 (0.06) \\
T & 0.05 (\textless 0.05) & 0.06 (\textless 0.05) & 0.05 (\textless 0.05) &  0.00 (0.73) & 0.01 (0.32) & 0.01 (0.43) \\
\end{tabular}
\end{ruledtabular}
\end{table*}

If the calibration data is well-suited to the problem of model selection, lower calibration RMSEs result in better agreement between film doses and planned or measured doses and, consequently, lower values of $\overline{\gamma}$ and higher $\gamma_{<1}$ in gamma analysis.  In Fig.~\ref{fig:correlation} it is shown that calibration RMSEs and $\overline{\gamma}$ or $\gamma_{<1}$ from Table~\ref{tab:Main} are significantly correlated.  A consequence of this is shown in Table~\ref{tab:Pair_models}. It compares the models of channel-independent perturbations employing paired difference test. Models are paired for each dosimetry protocol. Differences are in $\overline{\gamma}$ values from Table~\ref{tab:Main}. Mean differences between models are shown with dosimetry protocols grouped into protocols R and T. Protocols t were not used in this analysis in order to have both transmission mode and reflection mode scanning protocols equally weighted. The TN model provided significantly ($p < 0.05$) better results than the rest of models  bringing together protocols R and T. Observing R protocols alone, the MM model was found significantly worse than WM and TN. Observing T protocols alone, the WM model was found significantly worse than the rest of models. Including in the analysis the rest of results, the TN model provided the best results both for R and T protocols, the UD model provided better results than the MM model, for R protocols the WM model provided better results than MM and UD models, however, for T protocols the WM model provided the worst results.

The WM model neglects correlations between channels. Poor performance of the WM model with T protocols can signify that these correlations ({\it e.g.}, due to variations in the thickness of the active layer) are important and can not be neglected. Still, this outcome does not mean that MM or UD models are preferable to WM in transmission mode scanning, it depends on the uncertainty $\sigma_{D}$ (Eq.(\ref{uncertainty})), {\it i.e.}, on the properties of the dosimetry system under study. In fact, after analysing a sample of points from different test cases, $\sigma_{D}$ was found to be the reason why MM and UD models provided worse results than the WM model with R protocols. To illustrate the importance of $\sigma_{D}$, the protocol PR2p was calibrated with MM using only red and green color channels. In our dosimetry system the sensitometric curves of both channels were very similar and this caused the RMSE of the calibration to be 3300 cGy. From Eq.(\ref{MMuncertainty}), it follows that the accuracy of the MM model depends on the properties of the sensitometric curves, and can result in unacceptable uncertainties. Another flaw of the MM model is that all $\sigma_{k}$ are considered equal. This hypothesis is usually false. As an example, it was found for protocol PR3p:  $\sigma_{R} = 3.5$ cGy, $\sigma_{G} = 2.8$ cGy and $\sigma_{B} = 6.8$ cGy. The UD model corrects this deficiency, which could explain why it provided better results than the MM model. However, the accuracy of the UD model still depends on the properties of the sensitometric curves and of $\sigma_{k}$. With respect to the TN model, even though it is also submitted to Eq.(\ref{uncertainty}), it can be considered as a metamodel that minimizes $\sigma_{D}$ and can derive (as a limit) into the WM model, or into an intermediate case between models WM and UD. As a conclusion, we believe the superior performance of the TN model of channel-independent perturbations makes it the best choice for multichannel dosimetry.

Once the TN model was selected, gamma analysis was conducted for Pm and Mp protocols. Values of $\overline{\gamma}$ and $\gamma_{<1}$ for this protocols are shown in Table~\ref{tab:Main2}. RMSEs from the calibrations are not included since they are already present in Table~\ref{tab:Main} ({\it i.e.}, the RMSE from the calibration is independent of the evaluation dose distribution used for gamma analysis).

\subsection{Selection of dosimetry protocol}

\begin{table*}
\caption{\label{tab:Main2} Comparison of film doses ($d(r)$) with planned or measured doses ($D(r)$). Data are aggregated by model of channel-independent perturbations and dosimetry protocol. It completes Table~\ref{tab:Main} for the truncated normal distribution (TN) model.}
\begin{ruledtabular}
\begin{tabular}{l cc}
 & \multicolumn{2}{c}{TN}\\
\cline{2-3} 

& $\overline{\gamma}$ & $\gamma_{<1}$ \\
\hline
PR2m & 0.15 & 99.4 \\
Pt2m & 0.17 & 99.2 \\
PT2m & 0.20 & 98.5 \\
PR3m & 0.15 & 98.8 \\
Pt3m & 0.19 & 98.3 \\
PT3m & 0.18 & 98.4 \\
MR2p & 0.47 & 88.6 \\
Mt2p & 0.35 & 95.5 \\
MT2p & 0.43 & 91.2 \\
MR3p & 0.41 & 89.6 \\
Mt3p & 0.39 & 91.1 \\
MT3p & 0.33 & 95.2 \\
\end{tabular}
\end{ruledtabular}
\end{table*}

Comparisons of elements of the dosimetry process were made employing paired difference tests for the TN model. For each point of each test case the difference in $\gamma$ values between two protocols was calculated. Between both protocols, only one element of the dosimetry process was modified. Since some test cases were not present in some protocols ({\it e.g.}, Test H in protocols M), the numbers of test cases differ between the comparisons. Results of the comparisons are shown in Table~\ref{tab:Pair_main}.

\begin{table*}
\caption{\label{tab:Pair_main} Comparing dosimetry protocols. When protocols with trasmission-mode scanning were grouped together they are symbolized with: (t+T). Differences are in $\gamma$ values. Mean differences between protocols as well as the mean of the means and its standard deviation are shown. Negative values indicate that the first model obtained better results than the second one, and the opposite for positive values. Between parentheses, the $p$-value of the t-test is included. The null hypothesis is that the mean difference between paired observations is zero.}

\begin{subtable}{\linewidth}
\caption{Protocols t versus T.} \label{tab:Pair_tT}    
\begin{ruledtabular}
\begin{tabular}{ccccccccc}
P2p & P3p & M2m & M3m & P2m & P3m & M2p & M3p & Mean ($p$-value) \\
\hline
-0.034 & 0.013 & -0.002 & -0.046 & 0.032 & -0.007 & -0.132 & 0.023 & -0.02 $\pm$ 0.06 ($p = 0.33$) \\
\end{tabular}
\end{ruledtabular}
\end{subtable}

\begin{subtable}{\linewidth}
\caption{Protocols R versus T.} \label{tab:Pair_RT}     
\begin{ruledtabular}
\begin{tabular}{ccccccccc}
P2p & P3p & M2m & M3m & P2m & P3m & M2p & M3p & Mean ($p$-value) \\
\hline
0.005 & 0.046 & 0.020 & 0.021 & -0.003 & -0.019 & 0.071 & 0.168 & 0.04 $\pm$ 0.06 ($p = 0.11$) \\ 
\end{tabular}
\end{ruledtabular}
\end{subtable}

\begin{subtable}{\linewidth}
\caption{Protocols 2 versus 3.} \label{tab:Pair_23}      
\begin{ruledtabular}
\begin{tabular}{ccccccccc}
PRp & P(t+T)p & MRm & M(t+T)m & PRm & P(t+T)m & MRp & M(t+T)p & Mean ($p$-value) \\
\hline
0.038 & 0.061 & -0.066 & -0.003 & 0.008 & 0.024 & -0.064 & -0.001 & -0.001 $\pm$ 0.050 ($p = 0.97$) \\  
\end{tabular}
\end{ruledtabular}
\end{subtable}

\begin{subtable}{\linewidth}
\caption{Protocols p versus m.} \label{tab:Pair_em}      
\begin{ruledtabular}
\begin{tabular}{ccccccccc}
PR2 & P(t+T)2 & PR3 & P(t+T)3 & MR2 & M(t+T)2 & MR3 & M(t+T)3 & Mean ($p$-value) \\
\hline
0.084 & 0.027 & 0.049 & 0.002 & 0.232 & 0.102 & 0.179 & 0.085 & 0.10 $\pm$ 0.08 ($p < 0.05$) \\  
\end{tabular}
\end{ruledtabular}
\end{subtable}

\begin{subtable}{\linewidth}
\caption{Protocols P versus M.} \label{tab:Pair_EM}       
\begin{ruledtabular}
\begin{tabular}{ccccccccc}
R2p & (t+T)2p & R3p & (t+T)3p & R2m & (t+T)2m & R3m & (t+T)3m &  Mean ($p$-value) \\
\hline
-0.370 & -0.251 & -0.370 & -0.239 & -0.193 & -0.149 & -0.226 & -0.151 & -0.24 $\pm$ 0.09 ($p < 0.05$) \\ 
\end{tabular}
\end{ruledtabular}
\end{subtable}

\end{table*}

Table~\ref{tab:Pair_tT} and Table~\ref{tab:Pair_RT} compare transmission mode scanning with $75 \pm 5$ min time window (t) versus transmission mode scanning with $20 \pm 1$ h time window (T), and reflection mode scanning with $20 \pm 1$ h time window (R) versus transmission mode scanning with $20 \pm 1$ h time window, respectively. Protocols t provided better results than T, and T better than R. However, the differences are not significant. In Table~\ref{tab:Pair_23}, there is almost no difference between employing film type EBT2 (2) or EBT3 (3).
Table~\ref{tab:Pair_em} shows significant ($p < 0.05$) differences between calibration with planned (p) or measured (m) reference dose distributions. This result could be explained assuming that, for the EDW plan used in the calibration, the dose distribution measured with MatriXX has less uncertainty than the dose planned with Eclipse 10. Table~\ref{tab:Pair_EM} shows significantly ($p < 0.05$) better results for the gamma analysis with planned (P) than with measured (M) dose distributions. This is a consequence of the resolution of the evaluation dose distribution which is much worse for MatriXX. The resolution of the array affects negatively the value of the $\gamma$-index in spite of using bicubic interpolation. Swapping reference and evaluation dose distributions was discarded since it would induce noise artifacts\cite{Clasie:2012}.

Taking into account these comparisons, we selected the following dosimetry protocol: calibration with measured reference dose distributions, using film type EBT3, scanning in transmission mode with $75 \pm 5$ min time window and comparing the results with gamma analysis using planned evaluation dose distributions ({\it i.e.}, protocol Pt3m). Following this protocol allowed us to improve our previous $\gamma$-index tolerances from 4 \% 3 mm to 3 \% 3 mm or even 2.5 \% 2.5 mm, results are presented in Table~\ref{tab:final_gamma}.

\begin{table*}
\caption{\label{tab:final_gamma} Gamma analisys of the test cases with dosimetry model TN, protocol Pt3m and different tolerances: 4 \% 3 mm (with 20\% of the dose maximum ($D_{max}$) as threshold), 3 \% 3 mm (threshold 10\% of $D_{max}$) and 2.5 \% 2.5 mm (threshold 10\% of $D_{max}$).}

\begin{ruledtabular}
\begin{tabular}{l cc cc cc}
 & \multicolumn{2}{c}{$\gamma$ (4 \%, 3 mm)} & \multicolumn{2}{c}{$\gamma$ (3 \%, 3 mm)} & \multicolumn{2}{c}{$\gamma$ (2.5 \%, 2.5 mm)}\\
\cline{2-3} 
\cline{4-5}
\cline{6-7}

Test & $\overline{\gamma}$ & $\gamma_{<1}$ (\%) & $\overline{\gamma}$ & $\gamma_{<1}$ (\%) & $\overline{\gamma}$ & $\gamma_{<1}$ (\%) \\
\hline
A & 0.15 & 99.6 & 0.17 & 99.5 & 0.22 & 97.9\\
B & 0.23 & 97.7 & 0.26 & 96.0 & 0.34 & 93.6\\
C & 0.13 & 100  & 0.22 & 99.4 & 0.30 & 97.1\\
D & 0.19 & 96.7 & 0.13 & 99.9 & 0.18 & 99.3\\
E & 0.15 & 97.7 & 0.22 & 96.8 & 0.31 & 95.8\\
F & 0.14 & 100  & 0.14 & 99.8 & 0.18 & 99.3\\
G & 0.14 & 99.9 & 0.18 & 99.2 & 0.23 & 97.9\\
H & 0.18 & 98.6 & 0.22 & 97.7 & 0.31 & 94.8\\
I & 0.10 & 99.8 & 0.17 & 98.8 & 0.23 & 97.0\\
J & 0.15 & 99.5 & 0.20 & 98.3 & 0.27 & 96.1\\
K & 0.15 & 99.7 & 0.24 & 98.4 & 0.30 & 95.9\\
L & 0.17 & 99.7 & 0.24 & 98.0 & 0.33 & 93.5\\
M & 0.21 & 98.4 & 0.32 & 94.0 & 0.42 & 88.6\\
N & 0.39 & 93.5 & 0.52 & 83.5 & 0.66 & 71.9\\
\end{tabular}
\end{ruledtabular}
\end{table*}   

\subsection{Summary and recommendations}

With respect to the model of channel-independent perturbations:

\begin{enumerate}
\item We recommend using the truncated normal distribution model because it can be considered as a metamodel which minimizes the uncertainty in the dose inherent in the method of channel-independent perturbations. The weighted mean model neglects correlations between the channels, which can be important, and the accuracy of the Micke-Mayer model depends on the properties of the sensitometric curves, which can result in unacceptable uncertainties for particular dosimetry systems. Since the other models are either limit cases or particular cases of the TN model, the latter should provide at least as good results as them.
\item For film calibration using the TN model, it is recommended to calibrate each color channel first. After that, two parameters: $\tilde{\sigma}_{\Delta}$ and of $\sigma_{\Delta}$, are obtained optimizing the RMSE of the differences between film doses ($d(r)$) and reference doses ($D(r)$), according to Eq.(\ref{Dose}) and Eq.(\ref{eq_cal}).
\item Film doses can be calculated following a closed-form formula (Eq.(\ref{Dose})). In addition, the type B uncertainty in the dose implicit in the method can be calculated (Eq.(\ref{uncertainty})).
\end{enumerate}

With respect to the dosimetry protocol, and excluding the comparisons between the particular TPS and array dosimeter used in this study:

\begin{enumerate}
\item No significant differences were found between transmission and reflection mode scanning.
\item Short waiting-time windows can be employed without losing accuracy, as pointed out by Lewis \emph{et al}\cite{lewis:2012}.
\item No significant differences were found between using EBT2 or EBT3 films.
\end{enumerate}

\section{Conclusions}

Four models of channel-independent perturbations for multichannel film dosimetry were examined. Two of them based on the literature: a model which employs channel-independent perturbations as proposed by Micke \emph{et al}\cite{AMicke:2011} and further developed by Mayer \emph{et al}\cite{mayer:2012}, and another one which uses the weighted mean of all three channels to obtain the dose\cite{mendez:2013}. In addition to these, two novel models were proposed, a more realistic extension to the Micke-Mayer model which uses uniform distributed perturbations but allows the error terms to differ from one channel to another, and a truncated normal distribution, which comprises the other models as particular or limit cases.

A closed-form formula for dose calculation was derived for all four models, and it coincides with the published one\cite{mayer:2012} in the case of the Micke-Mayer model. In addition, Type B uncertainties in film dose due to the channel-independent perturbations method were obtained.

In order to assess the performance of the models, a set of tests was devised in which the dose distributions obtained from films were compared to either planned, or measured dose distributions. In these tests, the truncated normal distribution model provided the best agreement between film and reference doses, both for calibration and $\gamma$-index verification, and proved itself superior to both the weighted mean model, which neglects correlations between the channels, and the Micke-Mayer and the uniform distribution models, whose accuracy depends on the properties of the sensitometric curves. As a conclusion, we feel confident to recommend the truncated normal distribution model of channel-independent perturbations for multichannel dosimetry.
 
Along with the models, other factors which could influence the dosimetry process were also evaluated. No significant differences were found between transmission mode scanning and reflection mode scanning, between $75 \pm 5$ min versus $20 \pm 1$ h waiting-time window or between employing EBT2 or EBT3 films. However, significantly better results were obtained when a measured dose distribution was used instead of a planned one as reference for the calibration, and when a planned dose distribution was used instead of a measured one as evaluation for the $\gamma$-analysis.

\begin{acknowledgments}
The authors would like to thank Denis Brojan, V\'{i}ctor Hern\'{a}ndez and Sa\v{s}o Pulko for their contributions to this work.
\end{acknowledgments}

\providecommand{\noopsort}[1]{}\providecommand{\singleletter}[1]{#1}%


\begin{thebibliography}{34}%
\makeatletter
\providecommand \@ifxundefined [1]{%
 \@ifx{#1\undefined}
}%
\providecommand \@ifnum [1]{%
 \ifnum #1\expandafter \@firstoftwo
 \else \expandafter \@secondoftwo
 \fi
}%
\providecommand \@ifx [1]{%
 \ifx #1\expandafter \@firstoftwo
 \else \expandafter \@secondoftwo
 \fi
}%
\providecommand \natexlab [1]{#1}%
\providecommand \enquote  [1]{``#1''}%
\providecommand \bibnamefont  [1]{#1}%
\providecommand \bibfnamefont [1]{#1}%
\providecommand \citenamefont [1]{#1}%
\providecommand \href@noop [0]{\@secondoftwo}%
\providecommand \href [0]{\begingroup \@sanitize@url \@href}%
\providecommand \@href[1]{\@@startlink{#1}\@@href}%
\providecommand \@@href[1]{\endgroup#1\@@endlink}%
\providecommand \@sanitize@url [0]{\catcode `\\12\catcode `\$12\catcode
  `\&12\catcode `\#12\catcode `\^12\catcode `\_12\catcode `\%12\relax}%
\providecommand \@@startlink[1]{}%
\providecommand \@@endlink[0]{}%
\providecommand \url  [0]{\begingroup\@sanitize@url \@url }%
\providecommand \@url [1]{\endgroup\@href {#1}{\urlprefix }}%
\providecommand \urlprefix  [0]{URL }%
\providecommand \Eprint [0]{\href }%
\providecommand \doibase [0]{http://dx.doi.org/}%
\providecommand \selectlanguage [0]{\@gobble}%
\providecommand \bibinfo  [0]{\@secondoftwo}%
\providecommand \bibfield  [0]{\@secondoftwo}%
\providecommand \translation [1]{[#1]}%
\providecommand \BibitemOpen [0]{}%
\providecommand \bibitemStop [0]{}%
\providecommand \bibitemNoStop [0]{.\EOS\space}%
\providecommand \EOS [0]{\spacefactor3000\relax}%
\providecommand \BibitemShut  [1]{\csname bibitem#1\endcsname}%
\let\auto@bib@innerbib\@empty
\bibitem [{\citenamefont {Devic}\ \emph {et~al.}(2005)\citenamefont {Devic},
  \citenamefont {Seuntjens}, \citenamefont {Sham}, \citenamefont {Podgorsak},
  \citenamefont {Schmidtlein}, \citenamefont {Kirov},\ and\ \citenamefont
  {Soares}}]{devic:2005}%
  \BibitemOpen
  \bibfield  {author} {\bibinfo {author} {\bibfnamefont {S.}~\bibnamefont
  {Devic}}, \bibinfo {author} {\bibfnamefont {J.}~\bibnamefont {Seuntjens}},
  \bibinfo {author} {\bibfnamefont {E.}~\bibnamefont {Sham}}, \bibinfo {author}
  {\bibfnamefont {E.~B.}\ \bibnamefont {Podgorsak}}, \bibinfo {author}
  {\bibfnamefont {C.~R.}\ \bibnamefont {Schmidtlein}}, \bibinfo {author}
  {\bibfnamefont {A.~S.}\ \bibnamefont {Kirov}}, \ and\ \bibinfo {author}
  {\bibfnamefont {C.~G.}\ \bibnamefont {Soares}},\ }\bibfield  {title}
  {\enquote {\bibinfo {title} {Precise radiochromic film dosimetry using a
  flat-bed document scanner},}\ }\href {\doibase 10.1118/1.1929253} {\bibfield
  {journal} {\bibinfo  {journal} {Medical Physics}\ }\textbf {\bibinfo {volume}
  {32}},\ \bibinfo {pages} {2245--2253} (\bibinfo {year} {2005})}\BibitemShut
  {NoStop}%
\bibitem [{\citenamefont {Fuss}\ \emph {et~al.}(2007)\citenamefont {Fuss},
  \citenamefont {Sturtewagen}, \citenamefont {Wagter},\ and\ \citenamefont
  {Georg}}]{Fuss:2007}%
  \BibitemOpen
  \bibfield  {author} {\bibinfo {author} {\bibfnamefont {M.}~\bibnamefont
  {Fuss}}, \bibinfo {author} {\bibfnamefont {E.}~\bibnamefont {Sturtewagen}},
  \bibinfo {author} {\bibfnamefont {C.~D.}\ \bibnamefont {Wagter}}, \ and\
  \bibinfo {author} {\bibfnamefont {D.}~\bibnamefont {Georg}},\ }\bibfield
  {title} {\enquote {\bibinfo {title} {Dosimetric characterization of
  {GafChromic EBT} film and its implication on film dosimetry quality
  assurance},}\ }\href {http://stacks.iop.org/0031-9155/52/i=14/a=013}
  {\bibfield  {journal} {\bibinfo  {journal} {Physics in Medicine and Biology}\
  }\textbf {\bibinfo {volume} {52}},\ \bibinfo {pages} {4211} (\bibinfo {year}
  {2007})}\BibitemShut {NoStop}%
\bibitem [{\citenamefont {Paelinck}, \citenamefont {Neve},\ and\ \citenamefont
  {Wagter}(2007)}]{Paelinck:2007}%
  \BibitemOpen
  \bibfield  {author} {\bibinfo {author} {\bibfnamefont {L.}~\bibnamefont
  {Paelinck}}, \bibinfo {author} {\bibfnamefont {W.~D.}\ \bibnamefont {Neve}},
  \ and\ \bibinfo {author} {\bibfnamefont {C.~D.}\ \bibnamefont {Wagter}},\
  }\bibfield  {title} {\enquote {\bibinfo {title} {Precautions and strategies
  in using a commercial flatbed scanner for radiochromic film dosimetry},}\
  }\href {http://stacks.iop.org/0031-9155/52/i=1/a=015} {\bibfield  {journal}
  {\bibinfo  {journal} {Physics in Medicine and Biology}\ }\textbf {\bibinfo
  {volume} {52}},\ \bibinfo {pages} {231} (\bibinfo {year} {2007})}\BibitemShut
  {NoStop}%
\bibitem [{\citenamefont {van Battum}\ \emph {et~al.}(2008)\citenamefont {van
  Battum}, \citenamefont {Hoffmans}, \citenamefont {Piersma},\ and\
  \citenamefont {Heukelom}}]{battum:2008}%
  \BibitemOpen
  \bibfield  {author} {\bibinfo {author} {\bibfnamefont {L.~J.}\ \bibnamefont
  {van Battum}}, \bibinfo {author} {\bibfnamefont {D.}~\bibnamefont
  {Hoffmans}}, \bibinfo {author} {\bibfnamefont {H.}~\bibnamefont {Piersma}}, \
  and\ \bibinfo {author} {\bibfnamefont {S.}~\bibnamefont {Heukelom}},\
  }\bibfield  {title} {\enquote {\bibinfo {title} {Accurate dosimetry with
  {GafChromic EBT} film of a 6 {MV} photon beam in water: {What} level is
  achievable?}}\ }\href {\doibase 10.1118/1.2828196} {\bibfield  {journal}
  {\bibinfo  {journal} {Medical Physics}\ }\textbf {\bibinfo {volume} {35}},\
  \bibinfo {pages} {704--716} (\bibinfo {year} {2008})}\BibitemShut {NoStop}%
\bibitem [{\citenamefont {Marti\v{s}\'{\i}kov\'{a}}, \citenamefont
  {Ackermann},\ and\ \citenamefont {J\"{a}kel}(2008)}]{Martisikova:2008}%
  \BibitemOpen
  \bibfield  {author} {\bibinfo {author} {\bibfnamefont {M.}~\bibnamefont
  {Marti\v{s}\'{\i}kov\'{a}}}, \bibinfo {author} {\bibfnamefont
  {B.}~\bibnamefont {Ackermann}}, \ and\ \bibinfo {author} {\bibfnamefont
  {O.}~\bibnamefont {J\"{a}kel}},\ }\bibfield  {title} {\enquote {\bibinfo
  {title} {Analysis of uncertainties in {Gafchromic EBT} film dosimetry of
  photon beams},}\ }\href {http://stacks.iop.org/0031-9155/53/i=24/a=001}
  {\bibfield  {journal} {\bibinfo  {journal} {Physics in Medicine and Biology}\
  }\textbf {\bibinfo {volume} {53}},\ \bibinfo {pages} {7013} (\bibinfo {year}
  {2008})}\BibitemShut {NoStop}%
\bibitem [{\citenamefont {Richley}\ \emph {et~al.}(2010)\citenamefont
  {Richley}, \citenamefont {John}, \citenamefont {Coomber},\ and\ \citenamefont
  {Fletcher}}]{Richley:2010}%
  \BibitemOpen
  \bibfield  {author} {\bibinfo {author} {\bibfnamefont {L.}~\bibnamefont
  {Richley}}, \bibinfo {author} {\bibfnamefont {A.~C.}\ \bibnamefont {John}},
  \bibinfo {author} {\bibfnamefont {H.}~\bibnamefont {Coomber}}, \ and\
  \bibinfo {author} {\bibfnamefont {S.}~\bibnamefont {Fletcher}},\ }\bibfield
  {title} {\enquote {\bibinfo {title} {Evaluation and optimization of the new
  {EBT2} radiochromic film dosimetry system for patient dose verification in
  radiotherapy},}\ }\href {http://stacks.iop.org/0031-9155/55/i=9/a=012}
  {\bibfield  {journal} {\bibinfo  {journal} {Physics in Medicine and Biology}\
  }\textbf {\bibinfo {volume} {55}},\ \bibinfo {pages} {2601} (\bibinfo {year}
  {2010})}\BibitemShut {NoStop}%
\bibitem [{\citenamefont {Devic}(2011)}]{Devic:2011}%
  \BibitemOpen
  \bibfield  {author} {\bibinfo {author} {\bibfnamefont {S.}~\bibnamefont
  {Devic}},\ }\bibfield  {title} {\enquote {\bibinfo {title} {Radiochromic film
  dosimetry: past, present, and future},}\ }\href {\doibase
  10.1016/j.ejmp.2010.10.001} {\bibfield  {journal} {\bibinfo  {journal}
  {Physica medica}\ }\textbf {\bibinfo {volume} {27}},\ \bibinfo {pages}
  {122--134} (\bibinfo {year} {2011})}\BibitemShut {NoStop}%
\bibitem [{\citenamefont {Crijns}\ \emph {et~al.}(2013)\citenamefont {Crijns},
  \citenamefont {Maes}, \citenamefont {van~der Heide},\ and\ \citenamefont {den
  Heuvel}}]{crijns:2102}%
  \BibitemOpen
  \bibfield  {author} {\bibinfo {author} {\bibfnamefont {W.}~\bibnamefont
  {Crijns}}, \bibinfo {author} {\bibfnamefont {F.}~\bibnamefont {Maes}},
  \bibinfo {author} {\bibfnamefont {U.~A.}\ \bibnamefont {van~der Heide}}, \
  and\ \bibinfo {author} {\bibfnamefont {F.~V.}\ \bibnamefont {den Heuvel}},\
  }\bibfield  {title} {\enquote {\bibinfo {title} {Calibrating page sized
  {Gafchromic} {EBT3} films},}\ }\href {\doibase 10.1118/1.4771960} {\bibfield
  {journal} {\bibinfo  {journal} {Medical Physics}\ }\textbf {\bibinfo {volume}
  {40}},\ \bibinfo {eid} {012102} (\bibinfo {year} {2013})}\BibitemShut
  {NoStop}%
\bibitem [{\citenamefont {Niroomand-Rad}\ \emph {et~al.}(1998)\citenamefont
  {Niroomand-Rad}, \citenamefont {Blackwell}, \citenamefont {Coursey},
  \citenamefont {Gall}, \citenamefont {Galvin}, \citenamefont {McLaughlin},
  \citenamefont {Meigooni}, \citenamefont {Nath}, \citenamefont {Rodgers},\
  and\ \citenamefont {Soares}}]{aapm:55}%
  \BibitemOpen
  \bibfield  {author} {\bibinfo {author} {\bibfnamefont {A.}~\bibnamefont
  {Niroomand-Rad}}, \bibinfo {author} {\bibfnamefont {C.~R.}\ \bibnamefont
  {Blackwell}}, \bibinfo {author} {\bibfnamefont {B.~M.}\ \bibnamefont
  {Coursey}}, \bibinfo {author} {\bibfnamefont {K.~P.}\ \bibnamefont {Gall}},
  \bibinfo {author} {\bibfnamefont {J.~M.}\ \bibnamefont {Galvin}}, \bibinfo
  {author} {\bibfnamefont {W.~L.}\ \bibnamefont {McLaughlin}}, \bibinfo
  {author} {\bibfnamefont {A.~S.}\ \bibnamefont {Meigooni}}, \bibinfo {author}
  {\bibfnamefont {R.}~\bibnamefont {Nath}}, \bibinfo {author} {\bibfnamefont
  {J.~E.}\ \bibnamefont {Rodgers}}, \ and\ \bibinfo {author} {\bibfnamefont
  {C.~G.}\ \bibnamefont {Soares}},\ }\bibfield  {title} {\enquote {\bibinfo
  {title} {Radiochromic film dosimetry: {Recommendations of AAPM Radiation
  Therapy Committee Task Group} 55},}\ }\href {\doibase 10.1118/1.598407}
  {\bibfield  {journal} {\bibinfo  {journal} {Medical Physics}\ }\textbf
  {\bibinfo {volume} {25}},\ \bibinfo {pages} {2093--2115} (\bibinfo {year}
  {1998})}\BibitemShut {NoStop}%
\bibitem [{\citenamefont {Rink}, \citenamefont {Vitkin},\ and\ \citenamefont
  {Jaffray}(2007)}]{rink:2007}%
  \BibitemOpen
  \bibfield  {author} {\bibinfo {author} {\bibfnamefont {A.}~\bibnamefont
  {Rink}}, \bibinfo {author} {\bibfnamefont {I.~A.}\ \bibnamefont {Vitkin}}, \
  and\ \bibinfo {author} {\bibfnamefont {D.~A.}\ \bibnamefont {Jaffray}},\
  }\bibfield  {title} {\enquote {\bibinfo {title} {Energy dependence {(75 kVp
  to 18 MV)} of radiochromic films assessed using a real-time optical
  dosimeter},}\ }\href {\doibase 10.1118/1.2431425} {\bibfield  {journal}
  {\bibinfo  {journal} {Medical Physics}\ }\textbf {\bibinfo {volume} {34}},\
  \bibinfo {pages} {458--463} (\bibinfo {year} {2007})}\BibitemShut {NoStop}%
\bibitem [{\citenamefont {Richter}\ \emph {et~al.}(2009)\citenamefont
  {Richter}, \citenamefont {Pawelke}, \citenamefont {Karsch},\ and\
  \citenamefont {Woithe}}]{richter:2009}%
  \BibitemOpen
  \bibfield  {author} {\bibinfo {author} {\bibfnamefont {C.}~\bibnamefont
  {Richter}}, \bibinfo {author} {\bibfnamefont {J.}~\bibnamefont {Pawelke}},
  \bibinfo {author} {\bibfnamefont {L.}~\bibnamefont {Karsch}}, \ and\ \bibinfo
  {author} {\bibfnamefont {J.}~\bibnamefont {Woithe}},\ }\bibfield  {title}
  {\enquote {\bibinfo {title} {Energy dependence of {EBT-1} radiochromic film
  response for photon {(10 kVp--15 MVp)} and electron beams {(6--18 MeV)}
  readout by a flatbed scanner},}\ }\href {\doibase 10.1118/1.3253902}
  {\bibfield  {journal} {\bibinfo  {journal} {Medical Physics}\ }\textbf
  {\bibinfo {volume} {36}},\ \bibinfo {pages} {5506--5514} (\bibinfo {year}
  {2009})}\BibitemShut {NoStop}%
\bibitem [{\citenamefont {Arjomandy}\ \emph {et~al.}(2010)\citenamefont
  {Arjomandy}, \citenamefont {Tailor}, \citenamefont {Anand}, \citenamefont
  {Sahoo}, \citenamefont {Gillin}, \citenamefont {Prado},\ and\ \citenamefont
  {Vicic}}]{arjomandy:2010}%
  \BibitemOpen
  \bibfield  {author} {\bibinfo {author} {\bibfnamefont {B.}~\bibnamefont
  {Arjomandy}}, \bibinfo {author} {\bibfnamefont {R.}~\bibnamefont {Tailor}},
  \bibinfo {author} {\bibfnamefont {A.}~\bibnamefont {Anand}}, \bibinfo
  {author} {\bibfnamefont {N.}~\bibnamefont {Sahoo}}, \bibinfo {author}
  {\bibfnamefont {M.}~\bibnamefont {Gillin}}, \bibinfo {author} {\bibfnamefont
  {K.}~\bibnamefont {Prado}}, \ and\ \bibinfo {author} {\bibfnamefont
  {M.}~\bibnamefont {Vicic}},\ }\bibfield  {title} {\enquote {\bibinfo {title}
  {Energy dependence and dose response of {Gafchromic EBT2} film over a wide
  range of photon, electron, and proton beam energies},}\ }\href {\doibase
  10.1118/1.3373523} {\bibfield  {journal} {\bibinfo  {journal} {Medical
  Physics}\ }\textbf {\bibinfo {volume} {37}},\ \bibinfo {pages} {1942--1947}
  (\bibinfo {year} {2010})}\BibitemShut {NoStop}%
\bibitem [{\citenamefont {Lindsay}\ \emph {et~al.}(2010)\citenamefont
  {Lindsay}, \citenamefont {Rink}, \citenamefont {Ruschin},\ and\ \citenamefont
  {Jaffray}}]{lindsay:2010}%
  \BibitemOpen
  \bibfield  {author} {\bibinfo {author} {\bibfnamefont {P.}~\bibnamefont
  {Lindsay}}, \bibinfo {author} {\bibfnamefont {A.}~\bibnamefont {Rink}},
  \bibinfo {author} {\bibfnamefont {M.}~\bibnamefont {Ruschin}}, \ and\
  \bibinfo {author} {\bibfnamefont {D.}~\bibnamefont {Jaffray}},\ }\bibfield
  {title} {\enquote {\bibinfo {title} {Investigation of energy dependence of
  {EBT} and {EBT-2 Gafchromic} film},}\ }\href {\doibase 10.1118/1.3291622}
  {\bibfield  {journal} {\bibinfo  {journal} {Medical Physics}\ }\textbf
  {\bibinfo {volume} {37}},\ \bibinfo {pages} {571--576} (\bibinfo {year}
  {2010})}\BibitemShut {NoStop}%
\bibitem [{\citenamefont {Massillon-JL}\ \emph {et~al.}(2012)\citenamefont
  {Massillon-JL}, \citenamefont {Chiu-Tsao}, \citenamefont {Domingo-Munoz},\
  and\ \citenamefont {Chan}}]{massillon}%
  \BibitemOpen
  \bibfield  {author} {\bibinfo {author} {\bibfnamefont {G.}~\bibnamefont
  {Massillon-JL}}, \bibinfo {author} {\bibfnamefont {S.}~\bibnamefont
  {Chiu-Tsao}}, \bibinfo {author} {\bibfnamefont {I.}~\bibnamefont
  {Domingo-Munoz}}, \ and\ \bibinfo {author} {\bibfnamefont {M.}~\bibnamefont
  {Chan}},\ }\bibfield  {title} {\enquote {\bibinfo {title} {{Energy Dependence
  of the New Gafchromic EBT3 Film:Dose Response Curves for 50 KV, 6 and 15 MV
  X-Ray Beams}},}\ }\href {\doibase 10.4236/ijmpcero.2012.12008} {\bibfield
  {journal} {\bibinfo  {journal} {International Journal of Medical Physics,
  Clinical Engineering and Radiation Oncology}\ }\textbf {\bibinfo {volume}
  {1}},\ \bibinfo {pages} {60--65} (\bibinfo {year} {2012})}\BibitemShut
  {NoStop}%
\bibitem [{\citenamefont {Micke}, \citenamefont {Lewis},\ and\ \citenamefont
  {Yu}(2011)}]{AMicke:2011}%
  \BibitemOpen
  \bibfield  {author} {\bibinfo {author} {\bibfnamefont {A.}~\bibnamefont
  {Micke}}, \bibinfo {author} {\bibfnamefont {D.~F.}\ \bibnamefont {Lewis}}, \
  and\ \bibinfo {author} {\bibfnamefont {X.}~\bibnamefont {Yu}},\ }\bibfield
  {title} {\enquote {\bibinfo {title} {Multichannel film dosimetry with
  nonuniformity correction},}\ }\href {\doibase 10.1118/1.3576105} {\bibfield
  {journal} {\bibinfo  {journal} {Medical Physics}\ }\textbf {\bibinfo {volume}
  {38}},\ \bibinfo {pages} {2523--2534} (\bibinfo {year} {2011})}\BibitemShut
  {NoStop}%
\bibitem [{\citenamefont {van Hoof}\ \emph {et~al.}(2012)\citenamefont {van
  Hoof}, \citenamefont {Granton}, \citenamefont {Landry}, \citenamefont
  {Podesta},\ and\ \citenamefont {Verhaegen}}]{vanHoof:2012}%
  \BibitemOpen
  \bibfield  {author} {\bibinfo {author} {\bibfnamefont {S.~J.}\ \bibnamefont
  {van Hoof}}, \bibinfo {author} {\bibfnamefont {P.~V.}\ \bibnamefont
  {Granton}}, \bibinfo {author} {\bibfnamefont {G.}~\bibnamefont {Landry}},
  \bibinfo {author} {\bibfnamefont {M.}~\bibnamefont {Podesta}}, \ and\
  \bibinfo {author} {\bibfnamefont {F.}~\bibnamefont {Verhaegen}},\ }\bibfield
  {title} {\enquote {\bibinfo {title} {Evaluation of a novel triple-channel
  radiochromic film analysis procedure using {EBT2}},}\ }\href
  {http://stacks.iop.org/0031-9155/57/i=13/a=4353} {\bibfield  {journal}
  {\bibinfo  {journal} {Physics in Medicine and Biology}\ }\textbf {\bibinfo
  {volume} {57}},\ \bibinfo {pages} {4353} (\bibinfo {year}
  {2012})}\BibitemShut {NoStop}%
\bibitem [{\citenamefont {Mayer}\ \emph {et~al.}(2012)\citenamefont {Mayer},
  \citenamefont {Ma}, \citenamefont {Chen}, \citenamefont {Miller},
  \citenamefont {Belard}, \citenamefont {McDonough},\ and\ \citenamefont
  {O'Connell}}]{mayer:2012}%
  \BibitemOpen
  \bibfield  {author} {\bibinfo {author} {\bibfnamefont {R.~R.}\ \bibnamefont
  {Mayer}}, \bibinfo {author} {\bibfnamefont {F.}~\bibnamefont {Ma}}, \bibinfo
  {author} {\bibfnamefont {Y.}~\bibnamefont {Chen}}, \bibinfo {author}
  {\bibfnamefont {R.~I.}\ \bibnamefont {Miller}}, \bibinfo {author}
  {\bibfnamefont {A.}~\bibnamefont {Belard}}, \bibinfo {author} {\bibfnamefont
  {J.}~\bibnamefont {McDonough}}, \ and\ \bibinfo {author} {\bibfnamefont
  {J.~J.}\ \bibnamefont {O'Connell}},\ }\bibfield  {title} {\enquote {\bibinfo
  {title} {Enhanced dosimetry procedures and assessment for {EBT2} radiochromic
  film},}\ }\href {\doibase 10.1118/1.3694100} {\bibfield  {journal} {\bibinfo
  {journal} {Medical Physics}\ }\textbf {\bibinfo {volume} {39}},\ \bibinfo
  {pages} {2147--2155} (\bibinfo {year} {2012})}\BibitemShut {NoStop}%
\bibitem [{\citenamefont {M\'{e}ndez}\ \emph {et~al.}(2013)\citenamefont
  {M\'{e}ndez}, \citenamefont {Hartman}, \citenamefont {Hudej}, \citenamefont
  {Strojnik},\ and\ \citenamefont {Casar}}]{mendez:2013}%
  \BibitemOpen
  \bibfield  {author} {\bibinfo {author} {\bibfnamefont {I.}~\bibnamefont
  {M\'{e}ndez}}, \bibinfo {author} {\bibfnamefont {V.}~\bibnamefont {Hartman}},
  \bibinfo {author} {\bibfnamefont {R.}~\bibnamefont {Hudej}}, \bibinfo
  {author} {\bibfnamefont {A.}~\bibnamefont {Strojnik}}, \ and\ \bibinfo
  {author} {\bibfnamefont {B.}~\bibnamefont {Casar}},\ }\bibfield  {title}
  {\enquote {\bibinfo {title} {Gafchromic {EBT2} film dosimetry in reflection
  mode with a novel plan-based calibration method},}\ }\href {\doibase
  10.1118/1.4772075} {\bibfield  {journal} {\bibinfo  {journal} {Medical
  Physics}\ }\textbf {\bibinfo {volume} {40}},\ \bibinfo {eid} {011720}
  (\bibinfo {year} {2013})}\BibitemShut {NoStop}%
\bibitem [{\citenamefont {Reinhardt}\ \emph {et~al.}(2012)\citenamefont
  {Reinhardt}, \citenamefont {Hillbrand}, \citenamefont {Wilkens},\ and\
  \citenamefont {Assmann}}]{reinhardt:2012}%
  \BibitemOpen
  \bibfield  {author} {\bibinfo {author} {\bibfnamefont {S.}~\bibnamefont
  {Reinhardt}}, \bibinfo {author} {\bibfnamefont {M.}~\bibnamefont
  {Hillbrand}}, \bibinfo {author} {\bibfnamefont {J.~J.}\ \bibnamefont
  {Wilkens}}, \ and\ \bibinfo {author} {\bibfnamefont {W.}~\bibnamefont
  {Assmann}},\ }\bibfield  {title} {\enquote {\bibinfo {title} {Comparison of
  gafchromic {EBT2} and {EBT3} films for clinical photon and proton beams},}\
  }\href {\doibase 10.1118/1.4737890} {\bibfield  {journal} {\bibinfo
  {journal} {Medical Physics}\ }\textbf {\bibinfo {volume} {39}},\ \bibinfo
  {pages} {5257--5262} (\bibinfo {year} {2012})}\BibitemShut {NoStop}%
\bibitem [{\citenamefont {Park}\ \emph {et~al.}(2012)\citenamefont {Park},
  \citenamefont {Kang}, \citenamefont {Cheong}, \citenamefont {Hwang},
  \citenamefont {Kim}, \citenamefont {Han}, \citenamefont {Lee}, \citenamefont
  {Kim}, \citenamefont {Bae}, \citenamefont {Kim}, \citenamefont {Kim},
  \citenamefont {Oh},\ and\ \citenamefont {Suh}}]{park:2012}%
  \BibitemOpen
  \bibfield  {author} {\bibinfo {author} {\bibfnamefont {S.}~\bibnamefont
  {Park}}, \bibinfo {author} {\bibfnamefont {S.-K.}\ \bibnamefont {Kang}},
  \bibinfo {author} {\bibfnamefont {K.-H.}\ \bibnamefont {Cheong}}, \bibinfo
  {author} {\bibfnamefont {T.}~\bibnamefont {Hwang}}, \bibinfo {author}
  {\bibfnamefont {H.}~\bibnamefont {Kim}}, \bibinfo {author} {\bibfnamefont
  {T.}~\bibnamefont {Han}}, \bibinfo {author} {\bibfnamefont {M.-Y.}\
  \bibnamefont {Lee}}, \bibinfo {author} {\bibfnamefont {K.}~\bibnamefont
  {Kim}}, \bibinfo {author} {\bibfnamefont {H.}~\bibnamefont {Bae}}, \bibinfo
  {author} {\bibfnamefont {H.~S.}\ \bibnamefont {Kim}}, \bibinfo {author}
  {\bibfnamefont {J.~H.}\ \bibnamefont {Kim}}, \bibinfo {author} {\bibfnamefont
  {S.~J.}\ \bibnamefont {Oh}}, \ and\ \bibinfo {author} {\bibfnamefont {J.-S.}\
  \bibnamefont {Suh}},\ }\bibfield  {title} {\enquote {\bibinfo {title}
  {Variations in dose distribution and optical properties of {Gafchromic EBT2}
  film according to scanning mode},}\ }\href {\doibase 10.1118/1.3700731}
  {\bibfield  {journal} {\bibinfo  {journal} {Medical Physics}\ }\textbf
  {\bibinfo {volume} {39}},\ \bibinfo {pages} {2524--2535} (\bibinfo {year}
  {2012})}\BibitemShut {NoStop}%
\bibitem [{\citenamefont {Devic}\ \emph {et~al.}(2010)\citenamefont {Devic},
  \citenamefont {Aldelaijan}, \citenamefont {Mohammed}, \citenamefont {Tomic},
  \citenamefont {Liang}, \citenamefont {DeBlois},\ and\ \citenamefont
  {Seuntjens}}]{devic:2010}%
  \BibitemOpen
  \bibfield  {author} {\bibinfo {author} {\bibfnamefont {S.}~\bibnamefont
  {Devic}}, \bibinfo {author} {\bibfnamefont {S.}~\bibnamefont {Aldelaijan}},
  \bibinfo {author} {\bibfnamefont {H.}~\bibnamefont {Mohammed}}, \bibinfo
  {author} {\bibfnamefont {N.}~\bibnamefont {Tomic}}, \bibinfo {author}
  {\bibfnamefont {L.-H.}\ \bibnamefont {Liang}}, \bibinfo {author}
  {\bibfnamefont {F.}~\bibnamefont {DeBlois}}, \ and\ \bibinfo {author}
  {\bibfnamefont {J.}~\bibnamefont {Seuntjens}},\ }\bibfield  {title} {\enquote
  {\bibinfo {title} {Absorption spectra time evolution of {EBT-2 model
  GAFCHROMIC} film},}\ }\href {\doibase 10.1118/1.3378675} {\bibfield
  {journal} {\bibinfo  {journal} {Medical Physics}\ }\textbf {\bibinfo {volume}
  {37}},\ \bibinfo {pages} {2207--2214} (\bibinfo {year} {2010})}\BibitemShut
  {NoStop}%
\bibitem [{\citenamefont {Fiandra}\ \emph {et~al.}(2006)\citenamefont
  {Fiandra}, \citenamefont {Ricardi}, \citenamefont {Ragona}, \citenamefont
  {Anglesio}, \citenamefont {Giglioli}, \citenamefont {Calamia},\ and\
  \citenamefont {Lucio}}]{fiandra:2006}%
  \BibitemOpen
  \bibfield  {author} {\bibinfo {author} {\bibfnamefont {C.}~\bibnamefont
  {Fiandra}}, \bibinfo {author} {\bibfnamefont {U.}~\bibnamefont {Ricardi}},
  \bibinfo {author} {\bibfnamefont {R.}~\bibnamefont {Ragona}}, \bibinfo
  {author} {\bibfnamefont {S.}~\bibnamefont {Anglesio}}, \bibinfo {author}
  {\bibfnamefont {F.~R.}\ \bibnamefont {Giglioli}}, \bibinfo {author}
  {\bibfnamefont {E.}~\bibnamefont {Calamia}}, \ and\ \bibinfo {author}
  {\bibfnamefont {F.}~\bibnamefont {Lucio}},\ }\bibfield  {title} {\enquote
  {\bibinfo {title} {Clinical use of {EBT} model {Gafchromic} film in
  radiotherapy},}\ }\href {\doibase 10.1118/1.2362876} {\bibfield  {journal}
  {\bibinfo  {journal} {Medical Physics}\ }\textbf {\bibinfo {volume} {33}},\
  \bibinfo {pages} {4314--4319} (\bibinfo {year} {2006})}\BibitemShut {NoStop}%
\bibitem [{\citenamefont {Devic}\ \emph {et~al.}(2006)\citenamefont {Devic},
  \citenamefont {Wang}, \citenamefont {Tomic},\ and\ \citenamefont
  {Podgorsak}}]{devic:2006}%
  \BibitemOpen
  \bibfield  {author} {\bibinfo {author} {\bibfnamefont {S.}~\bibnamefont
  {Devic}}, \bibinfo {author} {\bibfnamefont {Y.-Z.}\ \bibnamefont {Wang}},
  \bibinfo {author} {\bibfnamefont {N.}~\bibnamefont {Tomic}}, \ and\ \bibinfo
  {author} {\bibfnamefont {E.~B.}\ \bibnamefont {Podgorsak}},\ }\bibfield
  {title} {\enquote {\bibinfo {title} {Sensitivity of linear {CCD} array based
  film scanners used for film dosimetry},}\ }\href {\doibase 10.1118/1.2357836}
  {\bibfield  {journal} {\bibinfo  {journal} {Medical Physics}\ }\textbf
  {\bibinfo {volume} {33}},\ \bibinfo {pages} {3993--3996} (\bibinfo {year}
  {2006})}\BibitemShut {NoStop}%
\bibitem [{\citenamefont {Lynch}\ \emph {et~al.}(2006)\citenamefont {Lynch},
  \citenamefont {Kozelka}, \citenamefont {Ranade}, \citenamefont {Li},
  \citenamefont {Simon},\ and\ \citenamefont {Dempsey}}]{lynch:2006}%
  \BibitemOpen
  \bibfield  {author} {\bibinfo {author} {\bibfnamefont {B.~D.}\ \bibnamefont
  {Lynch}}, \bibinfo {author} {\bibfnamefont {J.}~\bibnamefont {Kozelka}},
  \bibinfo {author} {\bibfnamefont {M.~K.}\ \bibnamefont {Ranade}}, \bibinfo
  {author} {\bibfnamefont {J.~G.}\ \bibnamefont {Li}}, \bibinfo {author}
  {\bibfnamefont {W.~E.}\ \bibnamefont {Simon}}, \ and\ \bibinfo {author}
  {\bibfnamefont {J.~F.}\ \bibnamefont {Dempsey}},\ }\bibfield  {title}
  {\enquote {\bibinfo {title} {Important considerations for radiochromic film
  dosimetry with flatbed {CCD} scanners and {EBT GAFCHROMIC} film},}\ }\href
  {\doibase 10.1118/1.2370505} {\bibfield  {journal} {\bibinfo  {journal}
  {Medical Physics}\ }\textbf {\bibinfo {volume} {33}},\ \bibinfo {pages}
  {4551--4556} (\bibinfo {year} {2006})}\BibitemShut {NoStop}%
\bibitem [{\citenamefont {Menegotti}, \citenamefont {Delana},\ and\
  \citenamefont {Martignano}(2008)}]{menegotti:2008}%
  \BibitemOpen
  \bibfield  {author} {\bibinfo {author} {\bibfnamefont {L.}~\bibnamefont
  {Menegotti}}, \bibinfo {author} {\bibfnamefont {A.}~\bibnamefont {Delana}}, \
  and\ \bibinfo {author} {\bibfnamefont {A.}~\bibnamefont {Martignano}},\
  }\bibfield  {title} {\enquote {\bibinfo {title} {Radiochromic film dosimetry
  with flatbed scanners: {A} fast and accurate method for dose calibration and
  uniformity correction with single film exposure},}\ }\href {\doibase
  10.1118/1.2936334} {\bibfield  {journal} {\bibinfo  {journal} {Medical
  Physics}\ }\textbf {\bibinfo {volume} {35}},\ \bibinfo {pages} {3078--3085}
  (\bibinfo {year} {2008})}\BibitemShut {NoStop}%
\bibitem [{\citenamefont {Saur}\ and\ \citenamefont
  {Frengen}(2008)}]{saur:2008}%
  \BibitemOpen
  \bibfield  {author} {\bibinfo {author} {\bibfnamefont {S.}~\bibnamefont
  {Saur}}\ and\ \bibinfo {author} {\bibfnamefont {J.}~\bibnamefont {Frengen}},\
  }\bibfield  {title} {\enquote {\bibinfo {title} {Gafchromic {EBT} film
  dosimetry with flatbed {CCD} scanner: {A} novel background correction method
  and full dose uncertainty analysis},}\ }\href {\doibase 10.1118/1.2938522}
  {\bibfield  {journal} {\bibinfo  {journal} {Medical Physics}\ }\textbf
  {\bibinfo {volume} {35}},\ \bibinfo {pages} {3094--3101} (\bibinfo {year}
  {2008})}\BibitemShut {NoStop}%
\bibitem [{\citenamefont {Ohuchi}(2007)}]{ohuchi:2007}%
  \BibitemOpen
  \bibfield  {author} {\bibinfo {author} {\bibfnamefont {H.}~\bibnamefont
  {Ohuchi}},\ }\bibfield  {title} {\enquote {\bibinfo {title} {High sensitivity
  radiochromic film dosimetry using an optical common-mode rejection and a
  reflective-mode flatbed color scanner},}\ }\href {\doibase 10.1118/1.2795828}
  {\bibfield  {journal} {\bibinfo  {journal} {Medical Physics}\ }\textbf
  {\bibinfo {volume} {34}},\ \bibinfo {pages} {4207--4212} (\bibinfo {year}
  {2007})}\BibitemShut {NoStop}%
\bibitem [{\citenamefont {JCGM}(2008)}]{JCGM:2008}%
  \BibitemOpen
  \bibfield  {author} {\bibinfo {author} {\bibnamefont {JCGM}},\ }\href@noop {}
  {\emph {\bibinfo {title} {JCGM 100:2008. Evaluation of measurement data —
  Guide to the expression of uncertainty in measurement}}},\ \bibinfo {edition}
  {1st}\ ed.\ (\bibinfo {year} {2008})\BibitemShut {NoStop}%
\bibitem [{\citenamefont {K\"{u}nzler}\ \emph {et~al.}(2009)\citenamefont
  {K\"{u}nzler}, \citenamefont {Fotina}, \citenamefont {Stock},\ and\
  \citenamefont {Georg}}]{Kunzler:2009}%
  \BibitemOpen
  \bibfield  {author} {\bibinfo {author} {\bibfnamefont {T.}~\bibnamefont
  {K\"{u}nzler}}, \bibinfo {author} {\bibfnamefont {I.}~\bibnamefont {Fotina}},
  \bibinfo {author} {\bibfnamefont {M.}~\bibnamefont {Stock}}, \ and\ \bibinfo
  {author} {\bibfnamefont {D.}~\bibnamefont {Georg}},\ }\bibfield  {title}
  {\enquote {\bibinfo {title} {Experimental verification of a commercial {Monte
  Carlo}-based dose calculation module for high-energy photon beams},}\ }\href
  {http://stacks.iop.org/0031-9155/54/i=24/a=008} {\bibfield  {journal}
  {\bibinfo  {journal} {Physics in Medicine and Biology}\ }\textbf {\bibinfo
  {volume} {54}},\ \bibinfo {pages} {7363} (\bibinfo {year}
  {2009})}\BibitemShut {NoStop}%
\bibitem [{\citenamefont {{R Core Team}}(2012)}]{R:software}%
  \BibitemOpen
  \bibfield  {author} {\bibinfo {author} {\bibnamefont {{R Core Team}}},\
  }\href {http://www.R-project.org/} {\emph {\bibinfo {title} {R: A Language
  and Environment for Statistical Computing}}},\ \bibinfo {organization} {R
  Foundation for Statistical Computing},\ \bibinfo {address} {Vienna, Austria}
  (\bibinfo {year} {2012}),\ \bibinfo {note} {{ISBN} 3-900051-07-0}\BibitemShut
  {NoStop}%
\bibitem [{\citenamefont {Van~Esch}\ \emph {et~al.}(2002)\citenamefont
  {Van~Esch}, \citenamefont {Bohsung}, \citenamefont {Sorvari}, \citenamefont
  {Tenhunen}, \citenamefont {Paiusco}, \citenamefont {Iori}, \citenamefont
  {Huyskens} \emph {et~al.}}]{vanesch:2002}%
  \BibitemOpen
  \bibfield  {author} {\bibinfo {author} {\bibfnamefont {A.}~\bibnamefont
  {Van~Esch}}, \bibinfo {author} {\bibfnamefont {J.}~\bibnamefont {Bohsung}},
  \bibinfo {author} {\bibfnamefont {P.}~\bibnamefont {Sorvari}}, \bibinfo
  {author} {\bibfnamefont {M.}~\bibnamefont {Tenhunen}}, \bibinfo {author}
  {\bibfnamefont {M.}~\bibnamefont {Paiusco}}, \bibinfo {author} {\bibfnamefont
  {M.}~\bibnamefont {Iori}}, \bibinfo {author} {\bibfnamefont {D.~P.}\
  \bibnamefont {Huyskens}},  \emph {et~al.},\ }\bibfield  {title} {\enquote
  {\bibinfo {title} {Acceptance tests and quality control ({QC}) procedures for
  the clinical implementation of intensity modulated radiotherapy ({IMRT})
  using inverse planning and the sliding window technique: experience from five
  radiotherapy departments},}\ }\href@noop {} {\bibfield  {journal} {\bibinfo
  {journal} {Radiotherapy and Oncology}\ }\textbf {\bibinfo {volume} {65}},\
  \bibinfo {pages} {53--70} (\bibinfo {year} {2002})}\BibitemShut {NoStop}%
\bibitem [{\citenamefont {Ju}\ \emph {et~al.}(2002)\citenamefont {Ju},
  \citenamefont {Ahn}, \citenamefont {Huh},\ and\ \citenamefont
  {Yeo}}]{ju:2002}%
  \BibitemOpen
  \bibfield  {author} {\bibinfo {author} {\bibfnamefont {S.~G.}\ \bibnamefont
  {Ju}}, \bibinfo {author} {\bibfnamefont {Y.~C.}\ \bibnamefont {Ahn}},
  \bibinfo {author} {\bibfnamefont {S.~J.}\ \bibnamefont {Huh}}, \ and\
  \bibinfo {author} {\bibfnamefont {I.~J.}\ \bibnamefont {Yeo}},\ }\bibfield
  {title} {\enquote {\bibinfo {title} {Film dosimetry for intensity modulated
  radiation therapy: Dosimetric evaluation},}\ }\href {\doibase
  10.1118/1.1449493} {\bibfield  {journal} {\bibinfo  {journal} {Medical
  Physics}\ }\textbf {\bibinfo {volume} {29}},\ \bibinfo {pages} {351--355}
  (\bibinfo {year} {2002})}\BibitemShut {NoStop}%
\bibitem [{\citenamefont {Clasie}\ \emph {et~al.}(2012)\citenamefont {Clasie},
  \citenamefont {Sharp}, \citenamefont {Seco}, \citenamefont {Flanz},\ and\
  \citenamefont {Kooy}}]{Clasie:2012}%
  \BibitemOpen
  \bibfield  {author} {\bibinfo {author} {\bibfnamefont {B.~M.}\ \bibnamefont
  {Clasie}}, \bibinfo {author} {\bibfnamefont {G.~C.}\ \bibnamefont {Sharp}},
  \bibinfo {author} {\bibfnamefont {J.}~\bibnamefont {Seco}}, \bibinfo {author}
  {\bibfnamefont {J.~B.}\ \bibnamefont {Flanz}}, \ and\ \bibinfo {author}
  {\bibfnamefont {H.~M.}\ \bibnamefont {Kooy}},\ }\bibfield  {title} {\enquote
  {\bibinfo {title} {Numerical solutions of the gamma-index in two and three
  dimensions},}\ }\href {http://stacks.iop.org/0031-9155/57/i=21/a=6981}
  {\bibfield  {journal} {\bibinfo  {journal} {Physics in Medicine and Biology}\
  }\textbf {\bibinfo {volume} {57}},\ \bibinfo {pages} {6981} (\bibinfo {year}
  {2012})}\BibitemShut {NoStop}%
\bibitem [{\citenamefont {Lewis}\ \emph {et~al.}(2012)\citenamefont {Lewis},
  \citenamefont {Micke}, \citenamefont {Yu},\ and\ \citenamefont
  {Chan}}]{lewis:2012}%
  \BibitemOpen
  \bibfield  {author} {\bibinfo {author} {\bibfnamefont {D.}~\bibnamefont
  {Lewis}}, \bibinfo {author} {\bibfnamefont {A.}~\bibnamefont {Micke}},
  \bibinfo {author} {\bibfnamefont {X.}~\bibnamefont {Yu}}, \ and\ \bibinfo
  {author} {\bibfnamefont {M.~F.}\ \bibnamefont {Chan}},\ }\bibfield  {title}
  {\enquote {\bibinfo {title} {An efficient protocol for radiochromic film
  dosimetry combining calibration and measurement in a single scan},}\ }\href
  {\doibase 10.1118/1.4754797} {\bibfield  {journal} {\bibinfo  {journal}
  {Medical Physics}\ }\textbf {\bibinfo {volume} {39}},\ \bibinfo {pages}
  {6339--6350} (\bibinfo {year} {2012})}\BibitemShut {NoStop}%
\end{thebibliography}
\end{document}